\documentclass[11pt,npb,preprint,showpacs,preprintnumbers,nofootinbib,amsmath]{revtex4}
\usepackage[dvips,final]{graphicx}
  \usepackage{amssymb}
   \usepackage{amsmath}
    \usepackage{epsfig}
     \usepackage{bm}
      \usepackage{pifont}

\usepackage{color}
\definecolor{mBlue}{rgb}{0,0,1}
 
\definecolor{mRed}{rgb}{1,0,0}
 
\definecolor{mGreen}{rgb}{0,1,0}

 \usepackage[dvips,final]{graphicx}
  \usepackage{amssymb,amsmath,epsfig,bm,pifont}
   \graphicspath{{./figs/}}

\newcommand{\gev}[1]{\relax\ifmmode{\text{GeV}^{#1}}               
                     \else{{GeV}$^{#1}${ }}\fi}                    

\begin{document}
\preprint{\hbox{RUB-TPII-04/2013}\\ }
\title{Theoretical description and measurement of the pion-photon
       transition form factor}
\thanks{Dedicated to the memory of our beloved friend and collaborator 
Alexander P. Bakulev.}
\author{S.~V.~Mikhailov}
\email{mikhs@theor.jinr.ru}
\affiliation{Bogoliubov Laboratory of Theoretical Physics, JINR,
             141980 Dubna, Russia \\} 
\author{A.~V.~Pimikov}
\email{pimikov@theor.jinr.ru}
\affiliation{Departamento de F\'{\i}sica Te\'orica -IFIC,
             Universidad de Valencia-CSIC, E-46100 Burjassot
             (Valencia), Spain 
             \\
             Bogoliubov Laboratory of Theoretical Physics, JINR,
             141980 Dubna, Russia \\} 
\author{N.~G.~Stefanis}
\email{stefanis@tp2.ruhr-uni-bochum.de}
\affiliation{Institut f\"{u}r Theoretische Physik II,
             Ruhr-Universit\"{a}t Bochum,
             D-44780 Bochum, Germany \\}

\begin{abstract}
Detailed predictions for the scaled pion-photon transition form factor
are given, derived with the method of light-cone sum rules and using
pion distribution amplitudes with two and three Gegenbauer
coefficients obtained from QCD sum rules with nonlocal condensates.
These predictions agree well with all experimental data that
are compatible with QCD scaling (and collinear factorization), but
disagree with the high-$Q^2$ data of the BaBar Collaboration that grow
with the momentum.
A good agreement of our predictions with results obtained from AdS/QCD
models and Dyson-Schwinger computations is found.
\end{abstract}

\maketitle

\section{Introduction}
\label{sec:intro}
The pion-photon transition form factor for two highly virtual photons,
$F^{\gamma^*\gamma^*\pi^0}(q_{1}^{2}, q_{2}^{2})$,
is considered to be a key hard exclusive process because it emerges as
a consequence of the factorization properties of
quantum chromodynamics (QCD).
All binding effects, governed by nonperturbative QCD interactions, are
isolated within the pion distribution amplitudes
$\varphi_{\pi}(x, \mu^2)$
(here $\mu^2$ is the normalization scale of order 1~GeV${}^2$)
for finding valence partons in the pion carrying a longitudinal
momentum fraction $x$ (struck quark) and
$\bar{x}\equiv 1-x$ (spectator) \cite{LB80,BL81-2phot}.
Assuming $q^2_{1}=-Q^2 \gg q_{2}^{2}=-q^2 > 0$, one can calculate
$F^{\gamma^*\gamma^*\pi^0}(Q^{2}, q^{2})$
in QCD perturbation theory including evolution \cite{ER80tmf,LB80}.

On the experimental side, the meson-photon transition form factors
can be measured using two-photon events is so-called single-tag
experiments of the generic form
$e^+ e^- \longrightarrow e^+ e^- M$,
where $M$ is a pseudoscalar meson $\pi^0, \eta, \eta'$,
in which only one electron (positron) is tagged, whereas the untagged
electron (positron) induces an uncertainty on the small virtuality
$|q^2|$ of the quasi-real photon.
However, such a photon has not a pointlike structure, rendering the
direct application of perturbative QCD unreliable.
Therefore, one has to employ another technique that allows one to
approach the real-photon limit in a more prudent way.
In our approach \cite{BMS02,BMS03,BMS05lat,BMPS11,BMPS12,SBMP12}, this
goal is achieved using the method of light-cone sum rules
\cite{Kho99,SY99,BBK89} --- see next section.
Our theoretical predictions, obtained this way, for the two-photon
process
$\gamma^* + \gamma \longrightarrow\pi^0 (\eta, \eta')$
are compared with the experimental data of various collaborations
\cite{CELLO91,CLEO98,BaBar09,Belle12} (pion case)
and \cite{CLEO98,BaBar11-BMS} ($\eta, \eta'$)
in Sec.\ \ref{sec:comp-ex-dat}.
The main characteristic of our predictions for
$Q^2F^{\gamma^*\gamma\pi^0}(Q^2, q^2\to 0)$ is a scaling behavior above
about $Q^2\simeq 10$~GeV$^2$, which is in accordance with the QCD
asymptotic prediction \cite{LB80}, both in trend and magnitude.
The auxetic \cite{SBMP12} behavior of the BaBar data for this process in
this region cannot be reproduced within collinear QCD.

\section{Theoretical description of the pion-photon
         transition form factor}
\label{sec:pi-ga-theory}
The calculation of the pion-photon transition form factor
\begin{equation}
  \int d^{4}z e^{-iq_{1}\cdot z}
  \langle
         \pi^0 (P)| T (j_{\mu}(z) j_{\nu}(0))| 0
  \rangle
=
  i\epsilon_{\mu\nu\alpha\beta}
  q_{1}^{\alpha} q_{2}^{\beta}
  F^{\gamma^{*}\gamma^{*}\pi^0}(Q^2,q^2)\ ,
\label{eq:matrix-element}
\end{equation}
where $j_\mu$ is the quark electromagnetic current,
is based on the following light-cone sum rule \cite{BBK89,Kho99}
\begin{eqnarray}
  Q^2 F^{\gamma^*\gamma^*\pi}\left(Q^2,q^2\right)
& = &
  \frac{\sqrt{2}}{3}f_\pi
  \left[
        \frac{Q^2}{m_{\rho}^2+q^2}
        \int_{x_{0}}^{1}
        \exp\left(
                  \frac{m_{\rho}^2-Q^2\bar{x}/x}{M^2}
            \right)
        \bar{\rho}(Q^2,x)
  \frac{dx}{x} \right.
\nonumber \\
&&
  + \left.
  \rule{0in}{0.25in}
      \int_{0}^{x_0} \bar{\rho}(Q^2,x)
        \frac{Q^2dx}{\bar{x}Q^2+x q^2}
  \right]
\, .
\label{eq:LCSR-FQq}
\end{eqnarray}
Here, the integration limits are defined by
$x_0=Q^2/\left(Q^2+s_0\right)$ and $s =\bar{x}Q^2/x$,
whereas $M^2$ is the Borel parameter and $m_\rho=0.77$~GeV denotes the
mass of the $\rho$ meson\footnote{In the actual calculation we
have also taken into account the width of the vector meson, as applied
in \cite{MS09}}.
The main ingredient in the above sum rule is the spectral density
$
 \bar{\rho}(Q^2,x)=(Q^2+s)\rho^\text{pert}(Q^2,s)
$,
where
\begin{eqnarray}
  \rho^\text{pert}(Q^2,s)
=
  \frac{1}{\pi} {\rm Im}F^{\gamma^*\gamma^*\pi^0}
  \left(Q^2,-s-i\varepsilon\right)
=
  \rho_\text{tw-2}
  +\rho_\text{tw-4}
  +\rho_\text{tw-6} .
\label{eq:rho-twists}
\end{eqnarray}
Each term corresponds to a definite twist contribution and is
calculable with the help of the analogous term of the hard-scattering
amplitude convoluted with the pion distribution amplitude of the same
twist.

In our analysis, presented in this note, we restrict our considerations
to the level of twist four, so that the transition form factor
acquires the form
\begin{eqnarray}
  F_{\gamma^*\gamma^*\pi^0}
\sim
  \left[
  T_\text{LO}
 +a_s(\mu^2) T_\text{NLO}
 +a_s^2(\mu^2) T_{\text{NNLO}_{\beta_0}}+\ldots
 \right]
\otimes
  \varphi_{\pi}^\text{(2)}(x,\mu^2)
  +\mathcal{O}\left(\frac{\delta^2}{Q^4}\right)~,
\label{eq:T_exp}
\end{eqnarray}
where the pion distribution amplitude (DA) $\varphi_{\pi}^\text{(2)}$
of twist two represents the nonperturbative part of the factorized
initial expression at the momentum scale
$\mu^2 \approx 1$~GeV$^2$ and $\delta^2$ sets the scale of the
twist-four term.
In our analysis we vary $\delta^2=0.19$~GeV$^2$,
(estimated in \cite{BMS02}),
in the range $[0.15-0.23]$~GeV$^2$, \cite{BMS03},
and use for the twist-four pion DA its asymptotic form
$\varphi_{\pi}^{(4)}(x,\mu^2))(80/3)\delta^2(\mu^2)x^2(1-x)^2$.
The meson DA can be expressed in terms of the coefficients
$a_n$ as an expansion over Gegenbauer harmonics, viz.,
($\xi = x-\bar{x}$)
\begin{equation}
 \label{G-expan}
  \varphi_{\pi}^{(2)}(x, \mu^2)
=
  6x\bar{x}
  \left(
       1 + \sum_{n=2,4, \ldots}^{\infty}a_n C_{n}^{3/2}(\xi)
  \right)
\end{equation}
with the normalization
$\int_{0}^{1} dx \varphi_{\pi}^{(2)}(x, \mu^2)=1$.
To compare with the experimental data at higher values $Q^2> \mu^2$, one
has to take into account evolution effects.
This is done here at the level of the next-to-leading-order (NLO)
Efremov-Radyushkin-Brodsky-Lepage (ERBL) evolution equation
\cite{ER80,LB80}.
In the next section, we will present predictions based on the
two Gegenbauer coefficients $a_2$ and $a_4$ determined long ago in
\cite{BMS01} via the moments
$\langle \xi^{N} \rangle_{\pi}
\equiv
  \int_{0}^{1} dx (2x-1)^{N} \varphi_{\pi}^{(2)}(x,\mu^2)$
with the help of QCD sum rules and nonlocal condensates
\cite{MR86,MR89,MR92}.
In addition, we will also show recent predictions \cite{BMPS11,SBMP12}
derived in a 3D analysis that makes use of the next higher Gegenbauer
coefficient $a_6$ that was found in \cite{BMS01} to be compatible with
zero but with large uncertainties.
The crucial characteristics of this type of distribution amplitudes
is that their endpoints are strongly suppressed.
This is related to the finite virtuality of the vacuum quarks
$\lambda_{q}^{2}=0.40(5)$~GeV$^2$
which implies that quarks at distances of the order of
$\Lambda \sim 1/\lambda \approx 0.31$~fm get correlated.
As a result, vacuum quarks with vanishing virtuality cannot migrate
into the pion bound state \cite{SSK99,SSK00}, because this is protected
by a mass gap.

\section{Comparing theory with experimental data}
\label{sec:comp-ex-dat}
The evaluation of the sum rule in Eq.\ (\ref{eq:LCSR-FQq}) is performed
with the following ingredients:
(i) We include in the spectral density the NLO radiative corrections
and also the twist four contribution.
(ii) The contributions related to the $\beta_0$ part of the
next-to-next-to-leading-order (NNLO) corrections \cite{MS09} and the
twist-six term, computed in \cite{ABOP10,ABOP12}, are included in the
form of systematic uncertainties.
This is justified, because for the values of the Borel parameter we
use around an average of $\overline{M^2}=0.75$~GeV$^2$, both
mentioned contributions are small, but enter with opposite signs so
that they partially cancel each other.
(Note that using a larger value $M^2=1.5$~GeV$^2$ as in \cite{ABOP10},
the twist-six term becomes negligible.)
(iii) The evolution is taken into account at the NLO using the
default scale setting
$\mu_{\rm R}^2=\mu_{\rm F}^2=Q^2$
and using the values
$\Lambda_{\rm QCD}^{(3)}=370$~MeV
and
$\Lambda_{\rm QCD}^{(4)}=304$~MeV,
consistent with
$\alpha_s(M_{Z}^2)=0.118$.

\begin{figure}[h]
 \centering
 \includegraphics[trim = 0mm 0mm 0mm 32 mm, clip, width=0.45\textwidth]{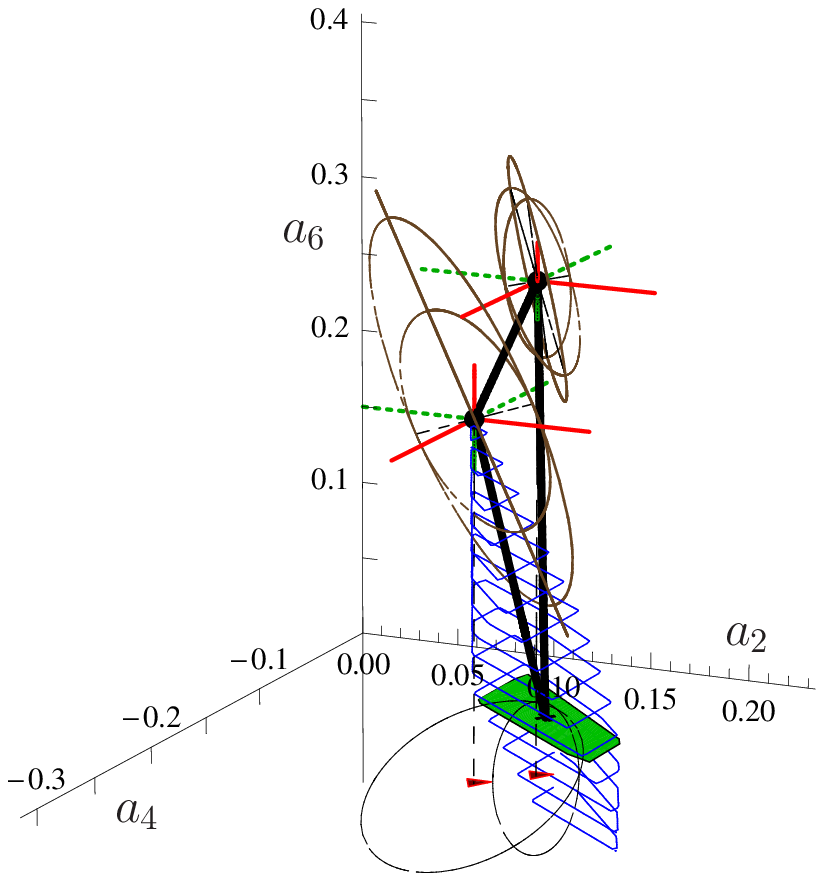}
 \hspace{-10mm}\hfill
 \includegraphics[width=0.57\textwidth]{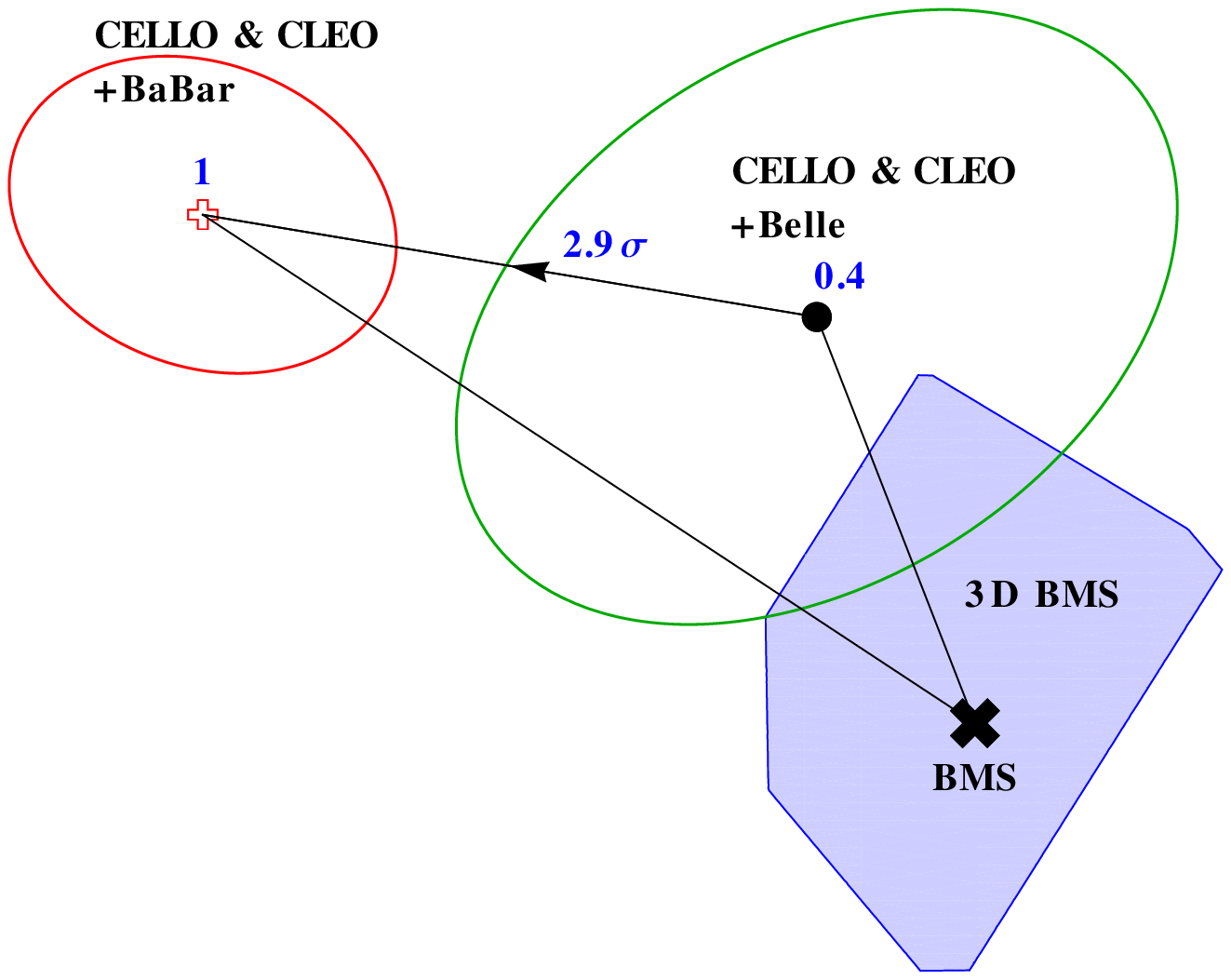}
\vspace*{3mm}
\caption{\label{fig:3D}
\textbf{Left panel:}
  The 3D confidence regions for the Gegenbauer coefficients
  ($a_2, a_4, a_6$) extracted from the analysis of two data sets:
  CELLO\&CLEO\&Belle (large left ellipsoid represented by its main ellipses)
  and CELLO\&CLEO\&BaBar (small ellipsoid) in comparison with the
  QCD SR result shown as a series of (blue) ``stairs'' along the $a_6$ axis.
  The shaded (green) rectangle, next to the projections of the
  3D ellipsoids, denotes the range of values of $a_2$ and $a_4$ determined
  by BMS via QCD sum rules with nonlocal condensates.
  The central points of the three 3D regions (large and small ellipsoids
  and BMS rectangle) build the vertices of a triangle.
  This triangle is also shown on the right panel of this figure.
\textbf{Right panel:}
  Quantification of the disagreement between the two data sets,
  termed (CELLO \& CLEO + Belle) vs. (CELLO \& CLEO +BaBar) in comparison
  with the BMS region of $(a_2, a_4$) values (shaded blue area).
  The 2D slice of the 3D figure, presented on the left, is
  shown in terms of a triangle with the values of the
  $\chi^2$-squared goodness of fit criterion
  $\chi^2/\text{ndf}$ marking its vertices. 
  }
\end{figure}
In Fig.\ \ref{fig:3D}, we present the results of our analysis for the
pion DAs, extracted from the experimental data, graphing them in terms
of the Gegenbauer coefficients $a_n$ in Eq.\ (\ref{G-expan}).
These coefficients, in both panels, have been obtained from the
statistical analysis of two data sets:
CELLO\&CLEO\&Belle (large left ellipsoid with the center named CCBe) and
CELLO\&CLEO\&BaBar (small ellipsoid with the center named CCBa).
The left panel of this figure shows the 3D confidence regions of the
coefficients ($a_2, a_4, a_6$) as a continuous stack of rectangles,
while the right panel reproduces a 2D slice of this 3D illustration.
The vertices of the triangle on the left panel are located at the
two best fit points, CCBe and CCBa, determined in the 3D analysis of the
corresponding data sets, while the third point {\footnotesize\ding{54}}
in the plane ($a_2, a_4$) denotes the BMS pion DA, derived from
QCD sum rules with nonlocal condensates in \cite{BMS01}.
Thus, the single (cut) plane shown on the right panel illustrates the
results of the statistical analysis of the data from the viewpoint
of the theoretical predictions.
In this figure we use the $\chi^2$-square goodness of fit criterion
$\chi^2/\text{ndf}$ (with ndf = number of degrees of freedom)
to mark the vertices of the triangle (right panel),
CCBa -- $\chi^2/\text{ndf}\gtrsim 1$
and CCBe -- $\chi^2/\text{ndf} \approx 0.4$.
We indicate the level of disagreement between the two considered
data sets quantitatively by means of the one standard deviation from the
CELLO\&CLEO\&Belle best-fit point that reaches for the
CCBa point the value $2.9\sigma_\text{\tiny CCBe}$
(more technical details can be found in \cite{SBMP12}).
As one sees from this figure, a considerable part of the 3D
BMS region --- blue rectangle on the right panel --- overlaps with the
$1\sigma_\text{\tiny CCBe}$ region of the CELLO\&CLEO\&Belle data.

\begin{figure}[h]
 \centering
 \includegraphics[width=0.46\textwidth]{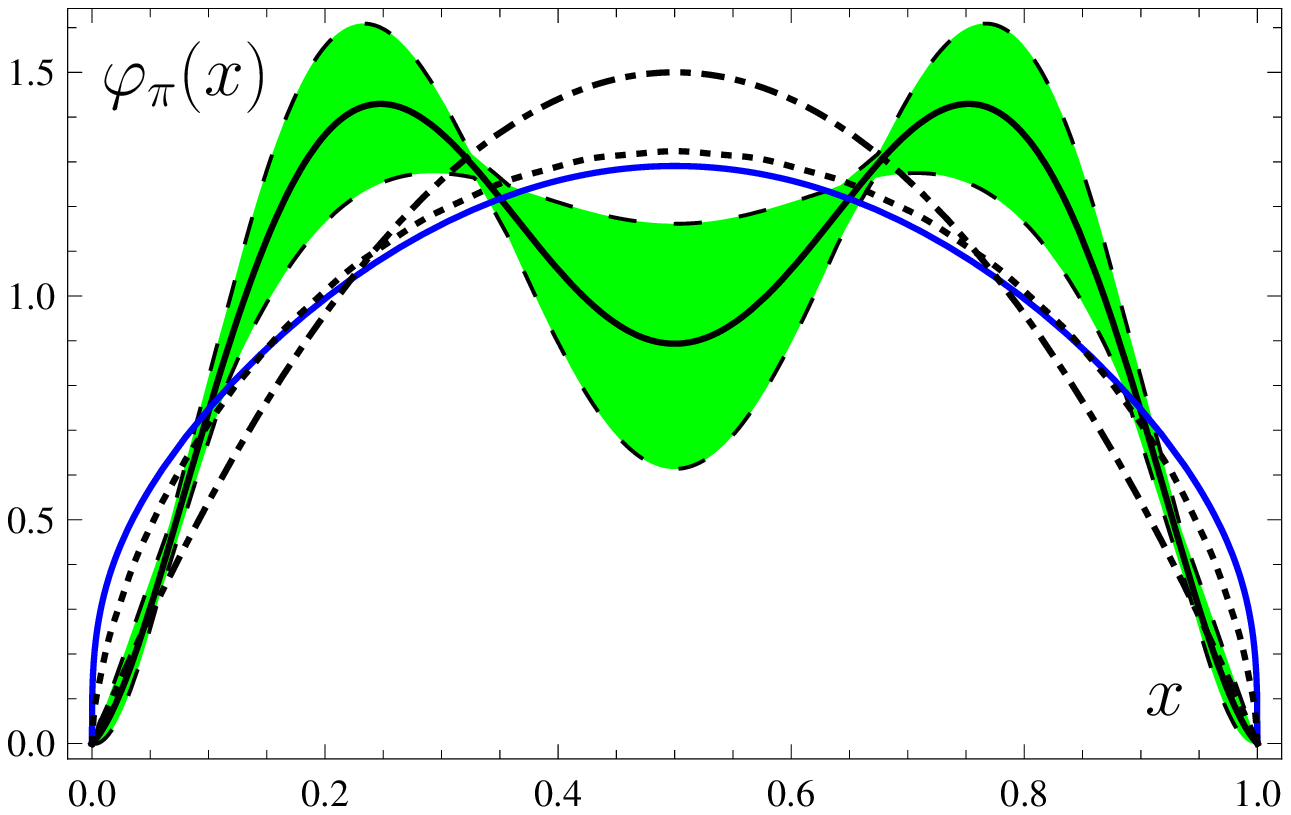}
 \hfill
 \includegraphics[width=0.47\textwidth]{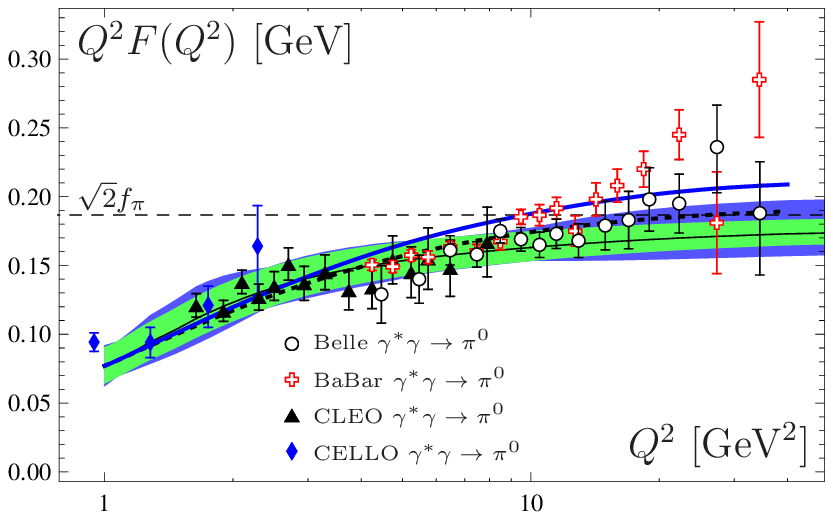}
 \vspace*{3mm}
\caption{\label{fig:predictions}
\textbf{Left panel}: Profiles of various pion DAs at the
normalization scale 2~GeV. 
The shaded band shows the ``bunch'' of the BMS DAs from \cite{BMS01},
with the solid line inside denoting the BMS model DA.
The dotted line is the model obtained in \cite{BCT11} from
AdS/QCD, whereas the (blue) solid line represents the result derived
in \cite{Chang:2013pq} using an approach based on the Dyson-Schwinger
equation.
For the sake of comparison, the asymptotic DA is also displayed ---
dashed-dotted line.
\textbf{Right panel}: Predictions for the scaled pion-photon
transition form factor in the momentum range covered by several
experiments.
The inner shaded band (in green) graphs our results for the BMS
``bunch'' shown on the left panel.
The upper and lower narrower strips (in blue color) denote
the additional areas of values emerging from the inclusion of the
coefficient $a_6$ --- 3D analysis.
The dotted line crossing the upper narrow strip represents the result
of the AdS/QCD model \cite{BCT11}, while the analogous result of the
DSE approach \cite{Chang:2013pq} is marked by the solid (blue) line
just above the upper narrow strip.
Note our further explanations in the text.}
\end{figure}

Let us now compare the ingredients of our analysis and its
predictions with other recent theoretical approaches.
This comparison will be selective.
Some examples of approaches (among many others) that can reproduce the
growing trend of the BaBar data can be found in
\cite{KV09,NV12epj,DK13prd,Kro10sud,Li:2013xna}, while for detailed
comparisons we refer to \cite{BMPS11,BMPS12}.
We show in the left panel of Fig.\ \ref{fig:predictions}
the profiles of the pion DAs in our scheme and two other
approaches.
The (green) shaded band between the broken lines contains the BMS-like
DAs from \cite{BMS01}, with the BMS model DA being denoted by a solid
line, evolved to the scale 2~GeV.
The two broad DAs show the profiles of, respectively,
an AdS/QCD model \cite{BCT11}, (reviewed in \cite{Brodsky:2013dca})
--- dotted curve ---
and the DA obtained within the Dyson-Schwinger-equation (DSE) based
approach of \cite{Chang:2013pq,Cloet:2013jya} --- (blue) solid curve.
The dashed-dotted line shows the asymptotic DA.
In the right panel of Fig.\ \ref{fig:predictions}, we show our
predictions for
$Q^2F_{\gamma^*\gamma^*\pi^0}(Q^2)$
in the form of a (green) shaded band that
contains the results of the 2D analysis, including the main
theoretical uncertainties.
The additional uncertainties, induced by the inclusion of the
coefficient $a_6$, are displayed in the form of (blue) strips
above and below the wider band.
For comparison, we also show the results for this observable
obtained from the most recent dynamical chiral-symmetry
preserving Dyson-Schwinger kernels currently available
\cite{Chang:2013pq} (blue solid line just above the band of our
predictions).
The result of the AdS/QCD model of \cite{BCT11} is displayed by a
dotted line grossly coinciding with our predictions.
In order to take into account the evolution with $Q^2$, we
applied the following procedure: (i) We expanded the
corresponding DAs in terms of Gegenbauer harmonics keeping only the
first 6 terms.
We justified this approximate treatment by computing the errors of
truncation.
We found that up to 40~GeV$^2$ the contributions from the last
considered term ($a_{12}$) are of the order of about 0.5\% for both
models AdS/QCD DA~\cite{BCT11} and DSE DA \cite{Chang:2013pq}.
These values confirm the saturation of the truncated Gegenbauer
expansion.
(ii) We evolved the expanded DAs from their proprietary scales
(1~GeV for the AdS/QCD DA~\cite{BCT11} and 2~GeV for the DSE DA
\cite{Chang:2013pq}) to the factorization scale which we set equal
to the large photon virtuality, i.e., $\mu_{\rm F}^2 = Q^2$.
(iii) We calculated the scaled form factors shown on the right panel of
Fig.\ \ref{fig:predictions} by means of the sum rule given by
Eq.\ (\ref{eq:LCSR-FQq}).
We emphasize that the predictions for these two external DA
models have been calculated in an approximate way serving mainly for
illustration.
Including a larger number of Gegenbauer harmonics into the expansion
of these DAs would eventually provide a still better agreement
with the asymptotic limit at high $Q^2$.
Nevertheless, as one observes, both curves agree rather well with
the extended band of our predictions and asymptotic QCD, though
the underlying pion DAs have very different profiles
at low-momentum scales --- Fig.\ \ref{fig:predictions}, left panel.
It is worth mentioning that our predictions show a similar
good agreement with those obtained in
\cite{ElBennich:2012ij,deMelo:2013zza}
within a QCD-inspired light-front quark approach.

\section{Conclusions}
\label{sec:concl}
We have discussed in this paper the theoretical calculation of
two-photon exclusive processes
$\gamma^* + \gamma\to M$ ($M=\pi^0, \eta, \eta'$)
using light-cone sum rules with perturbative input at the NLO,
combined with (nonperturbative) pion distribution amplitudes
obtained from QCD sum rules with nonlocal condensates.
We found that our predictions, which include various theoretical
uncertainties of the QCD approach,
conform with most available experimental data measured
by different Collaborations (CELLO, CLEO, BaBar, and Belle), apart
from the high-$Q^2$ data for the pion-photon transition form factor
of the BaBar experiment that violate QCD scaling above
$\sim 10$~GeV$^2$.
These predictions do not employ any fit parameters nor
do they need adjustment of the original parameters $a_2$ and $a_4$,
derived in \cite{BMS01} to the considered data sets.
The observed agreement improves further, if we include into
the Gegenbauer expansion of the pion distribution amplitude the third
coefficient $a_6$, promoting the original Bakulev-Mikhailov-Stefanis
approach \cite{BMS01} to a 3D analysis \cite{SBMP12}.
Remarkably, our predictions have large overlap with those derived from
holographic AdS/QCD \cite{BCT11} and the most recent DCS-improved
kernel in a DS-based approach \cite{Chang:2013pq}.
Our results support the conclusion \cite{SBMP12} that the distribution
amplitudes of the pion and the nonstrange component of $\eta$ and
$\eta'$, i.e.,
$|n\rangle=(1/\sqrt{2})(|u\bar{u}\rangle + |d\bar{d}\rangle)$,
have similar shapes with no indication for a significant
$SU(3)_{\rm F}$ flavor asymmetry.
In fact, our predictions for the scaled transition form factors
for both mesons, $\pi^0$ and $|n\rangle$, are in good agreement with
the scaling behavior at high $Q^2$, derived from the collinear
factorization properties of QCD.

\begin{acknowledgements}
Work partly supported by the Heisenberg-Landau Program 2013 and the
Russian Foundation for Fundamental Research
(Grants No.\ 12-02-00613a and 11-01-00182a).
A.V.P. acknowledges support by HadronPhysics2, Spanish Ministerio de
Economia y Competitividad
and EU FEDER under Contracts FPA2010-21750-C02-01, AIC10-D-000598, and
GVPrometeo2009/129.
\end{acknowledgements}



\newcommand{\noopsort}[1]{} \newcommand{\printfirst}[2]{#1}
  \newcommand{\singleletter}[1]{#1} \newcommand{\switchargs}[2]{#2#1}

\end{document}